\documentclass[preprint2]{aastex}

\newcommand{\lapprox} {\, \lower3pt\hbox{$\sim$}\llap{\raise2pt\hbox{$<$}}\,}
\newcommand{\gapprox} {\, \lower3pt\hbox{$\sim$}\llap{\raise2pt\hbox{$>$}}\,}

\begin{document}

\title{Electron-Electron Bremsstrahlung Emission and the Inference of Electron Flux Spectra in Solar Flares}

\author{Eduard P. Kontar\altaffilmark{1},  A. Gordon Emslie\altaffilmark{2}, Anna Maria Massone\altaffilmark{3},
    Michele Piana\altaffilmark{4}, John C. Brown\altaffilmark{1},
    and Marco Prato\altaffilmark{5}}

\altaffiltext{1}{Department of Physics \& Astronomy, University of
Glasgow G12 8QQ, UK \\ {eduard@astro.gla.ac.uk,
john@astro.gla.ac.uk}}

\altaffiltext{2}{Department of Physics, Oklahoma State University, Stillwater, OK 74078 \\
gordon.emslie@okstate.edu}

\altaffiltext{3}{CNR-INFM LAMIA, via Dodecaneso 33, I-16146
Genova, Italy(massone@ge.infm.it)}

\altaffiltext{4}{Dipartimento di Informatica, Universit\`a di
Verona, Ca\`\ Vignal 2, Strada le Grazie 15, I-37134 Verona, Italy
(michele.piana@univr.it)}

\altaffiltext{5}{Dipartimento di Matematica, Universit\`a di
Modena e Reggio Emilia, via Campi 213/b, I-41100 Modena, Italy
(marco.prato@unimo.it)}

\begin{abstract}

Although both electron-ion and electron-electron bremsstrahlung
contribute to the hard X-ray emission from solar flares, the
latter is normally ignored. Such an omission is not justified at
electron (and photon) energies above $\sim 300$~keV, and inclusion
of the additional electron-electron bremsstrahlung in general
makes the electron spectrum required to produce a given hard X-ray
spectrum steeper at high energies.

Unlike electron-ion bremsstrahlung, electron-electron
bremsstrahlung cannot produce photons of all energies up to the
maximum electron energy involved.  The maximum possible photon
energy depends on the angle between the direction of the emitting
electron and the emitted photon, and this suggests a diagnostic
for an upper cutoff energy and/or for the degree of beaming of the
accelerated electrons.

We analyze the large event of January 17, 2005 and show that the
upward break around 400~keV in the observed hard X-ray spectrum is
naturally accounted for by the inclusion of electron-electron
bremsstrahlung. Indeed, the mean source electron spectrum
recovered through a regularized inversion of the hard X-ray
spectrum, using a cross-section that includes both electron-ion
and electron-electron terms, has a relatively constant spectral
index $\delta$ over the range from electron kinetic energy $E =
200$~keV to $E = 1$~MeV. However, the level of detail discernible
in the recovered electron spectrum is not sufficient to determine
whether or not any upper cutoff energy exists.

\end{abstract}


\section{Introduction}

The spatially integrated hard X-ray spectrum $I(\epsilon)$
(photons~cm$^{-2}$~s$^{-1}$~keV$^{-1}$ at the Earth) is produced
by bremsstrahlung of accelerated electrons, characterized (Brown,
Emslie, \& Kontar 2003) by a {\it mean electron flux spectrum}
${\overline F}(E)$ (electrons~cm$^{-2}$~s$^{-1}$~keV$^{-1}$ at the
Sun) and related to $I(\epsilon)$ through

\begin{equation}
I(\epsilon) = {1 \over 4\pi R^2} \, \, {\bar n} V
\int_\epsilon^\infty  \overline F(E) \, Q(\epsilon,E) \, dE,
\label{def}
\end{equation}
where $Q(\epsilon,E)$ is the bremsstrahlung cross-section
(cm$^2$~keV$^{-1}$) differential in photon energy, $R = 1$~AU, and
the mean target density $\bar n$ (cm$^{-3}$) is defined by $\bar n
= V^{-1} \int n({\bf r}) \, dV$.  Bremsstrahlung in the energy
range $\gapprox 10$~keV is produced by energetic electrons
scattering off both protons/ions and electrons (both free and bound in atoms); these
contributions are summed to give the total differential bremsstrahlung
cross-section $Q(\epsilon,E)$.

For electron energies $\lapprox 300$~keV, the contribution from
electron-electron bremsstrahlung can be safely ignored (Haug
1975). However, for higher energies this is no longer the case.
Generally, for a given electron spectrum, the additional
electron-electron bremsstrahlung yield acts to flatten (harden)
the photon spectrum in this energy range (see, e.g., Haug 1975).
Equivalently, the inclusion of electron-electron bremsstrahlung
requires, for a given photon yield, a softer (steeper) electron
spectrum than would be required assuming electron-ion
bremsstrahlung alone.

The {\it Ramaty High Energy Solar Spectroscopic Imager} ({\em
RHESSI}) has opened a new era in the study of hard X-ray spectra
from solar flares.  With the high-resolution hard X-ray spectra
made available by {\it RHESSI}, an investigation into the form of
the hard X-ray spectrum above $\sim 300$~keV with the full
(electron-ion + electron-electron) cross-section is now warranted.

In this paper we therefore study the effect of adding the
electron-electron bremsstrahlung term on the mean
electron spectrum corresponding to a given hard X-ray spectrum.  In
\S2 we discuss the forms of the electron-ion and electron-electron
cross-sections, and we show that the presence of an upper limit to
the photon energy in the electron-electron process can in
principle provide evidence for a high-energy cutoff in the
electron spectrum and/or evidence of strong anisotropy in the
injected electron distribution.  In \S3 we discuss the sample
event (on January 17, 2005) chosen for analysis.  In \S4 we
present the form of the electron spectrum corresponding to the
observed photon spectrum, using both forward-fitting (e.g, Holman
et al. 2003) and regularization (Piana et al. 2003) techniques in
conjunction with the full (electron-ion + electron-electron)
bremsstrahlung cross-section. In \S5 we discuss the results
obtained, and in particular we point out that certain features in
the electron spectrum inferred using the electron-ion
bremsstrahlung cross-section are artifacts that can vanish when
the full, correct, cross-section is employed.

\section{Form of the Bremsstrahlung Cross-Section}

The cross-section (e.g., Koch \& Motz 1959) for electron-ion
bremsstrahlung scales as $Z^2$, where $Z$ is the atomic number of
the ion.  Further, in consideration of electron-electron
bremsstrahlung, the possible binding of target electrons to their
host ions in a neutral or partially-ionized medium is not
significant (E. Haug, personal communication). Hence, in a
quasi-neutral target of particles with atomic number $Z$, the
bremsstrahlung cross-section per atom for emission of a photon of
energy $\epsilon$ by an electron of energy $E$ is in general equal
to

\begin{equation}
Q (\epsilon, E) = Z^2 Q_{e-p} (\epsilon, E) + Z Q_{e-e} (\epsilon,
E). \label{qtot}
\end{equation}
were $Q_{e-p} (\epsilon, E)$ and $Q_{e-e} (\epsilon, E)$ are the
cross-sections for bremsstrahlung in electron-proton, and
electron-electron collisions, respectively.

The form of $Q_{e-e}(\epsilon,E)$, averaged over solid angle {\it
in the rest frame of the target electron} has been given\footnote
{Note that in the formula for $H(\epsilon,k,x)$ for the case $k >
{1 \over 2}$ (page 347 of that paper), the term $(\epsilon r /x +
s)$ on line 3 of the equation should be replaced by $(\epsilon r
/w + s)$ -- E. Haug, personal communication.} by Haug (1998),
while the solid-angle-averaged form in the {\it zero-momentum}
(``center-of mass'') frame has been given by Haug (1989). Neither
of these formulae are strictly appropriate to the case of an
electron beam incident on a warm plasma: the target electrons,
unlike the ions, have a velocity that may be comparable to the
velocity of the electrons in the impinging beam, so that a range
of injected particle/target particle relative velocities exist for
a given injected electron energy. However, as verified through
numerical simulation, for low electron energies (from $\sim 10$ to
$\sim 200$~keV), the form of ${\overline F}(E)$ corresponding to a
given hard X-ray spectrum, inferred using the
cross-section~(\ref{qtot}), does not differ, within statistical
uncertainties in the photon flux, from the form of ${\overline
F}(E)$ obtained using the electron-ion cross-section alone. Hence,
only at electron energies $\gapprox 200$~keV is the inclusion of
the electron-electron bremsstrahlung term necessary, and, for such
energies the velocity of the target particles {\it is} relatively
insignificant.  The target particle rest frame is, therefore, a
better approximation to the observer frame than is the
zero-momentum frame.  Hence, use of the electron-electron
cross-section in the {\it target particle} rest frame is more
appropriate.

When the maximum electron energy is much larger than the photon
energies under consideration, the photon spectrum resulting from a
power-law spectrum of electrons ${\overline F}(E) \sim
E^{-\delta}$ is also close to power-law form $I(\epsilon) \sim
\epsilon^{-\gamma}$ (Haug 1989). However, while for pure
electron-ion bremsstrahlung $\gamma \simeq \delta+1$, for pure
electron-electron bremsstrahlung a significantly shallower photon
spectrum, with $\gamma \simeq \delta$, results.  Thus, the
importance of the electron-electron bremsstrahlung contribution
increases with photon energy and the enhanced emission per
electron leads to a flattening of the photon spectrum
$I(\epsilon)$ produced by a given ${\overline F}(E)$, or,
equivalently, a steepening of the ${\overline F}(E)$ form required
to produce a given $I(\epsilon)$.

\begin{figure}[pht]
\epsscale{0.99} \plotone{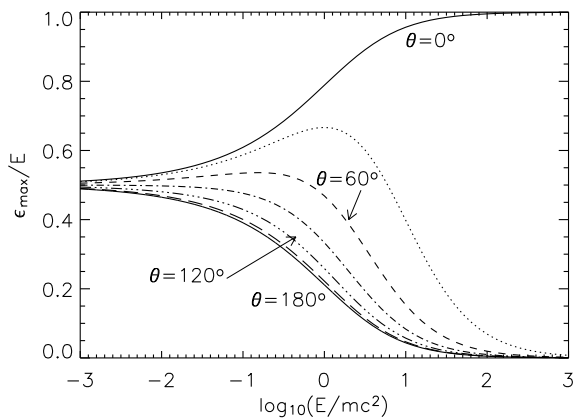} \caption{Maximum photon energy
$\epsilon_{\rm max}$ produced by electron-electron bremsstrahlung,
expressed as a fraction of the incident electron energy $E$ (in
units of the electron rest mass $mc^2$), for various values of
$\theta$, the angle between the incoming electron and the outgoing
photon trajectories For clarity, only curves for $\theta =
0^\circ, 60^\circ, 120^\circ,$ and $180^\circ$ are labelled; the
curves for $\theta = 30^\circ, 90^\circ$ and $150^\circ$ lie
between these.} \label{epsmax}
\end{figure}

It is also important to note that while the electron-ion
cross-section is finite for all $\epsilon < E$, the ``laboratory
frame'' cross-section for electron-electron bremsstrahlung
vanishes above a maximum photon energy, due to the necessarily
finite energy carried by the recoiling target electron.
Quantitatively (Haug 1975),
%\begin{equation} \epsilon _{\rm max} =mc^2 \, \frac{\psi}{1-\sqrt
%\psi \, \cos \theta}, \quad \psi=\frac E{E+2mc^2} \label{emax}
%\end{equation}

\begin{equation} \epsilon _{\rm max} = \frac{E}{E + 2 -\sqrt{E(E+2)}
\, \cos \theta},\label{emax}
\end{equation}
where $E$ is the electron kinetic energy in the laboratory frame (in units
of the electron rest energy $mc^2$) and $\theta$ is the angle
between the incoming electron and the outgoing photon
trajectories.  For highly {\it non}-relativistic electrons ($E \ll
1$), $\epsilon_{\rm max} \rightarrow E/2$ for all values of
$\theta$.  Only for highly relativistic electrons ($E \gg 1$) and
$\theta = 0$ (a singular case) does $\epsilon_{\rm max}
\rightarrow E$; for all other situations $\epsilon_{\rm max}$ is
less than $E$ and approaches $0$ as $E \rightarrow \infty$
(Figure~\ref{epsmax}).

This result has important implications for the form of the photon
spectrum produced by electron-electron bremsstrahlung.  If the
electron spectrum ${\overline F}(E)$ has a maximum energy $E_{\rm
max}$ that is not highly relativistic, then while electron-ion
bremsstrahlung will generate photons at all energies up to $E_{\rm
max}$, electron-electron bremsstrahlung will produce no photons at
all in the range $\epsilon_{\rm max} < \epsilon \le E_{\rm max}$.
The entire spectrum above $\epsilon_{\rm max}$ will therefore be
produced completely by electron-ion bremsstrahlung; the flattening
of the photon spectrum associated with the electron-electron
contribution disappears and the relationship between $I(\epsilon)$
and ${\bar F}(E)$ reverts to the form appropriate to electron-ion
bremsstrahlung alone.

\begin{figure}[pht]
\epsscale{1.0} \plottwo{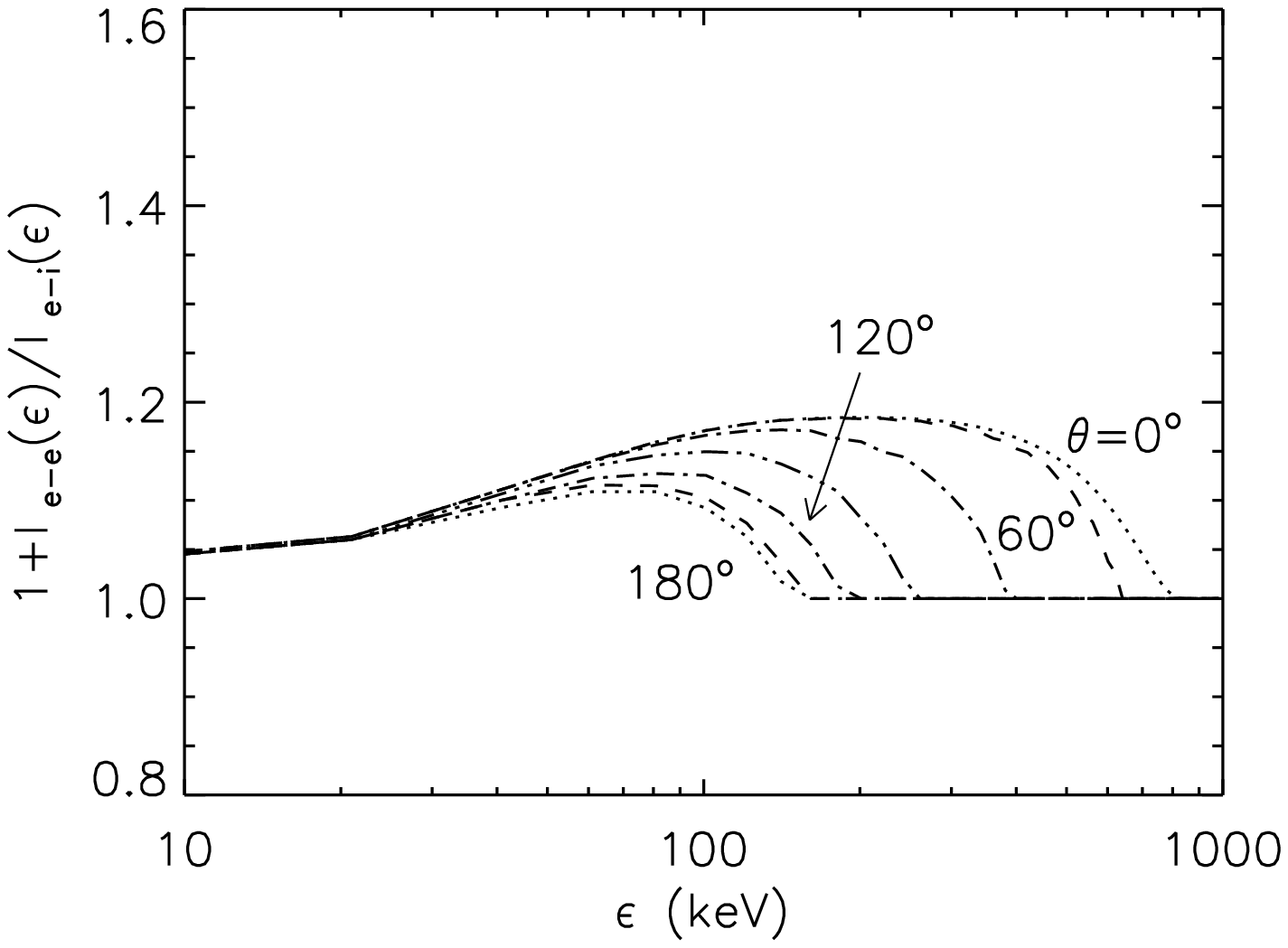}{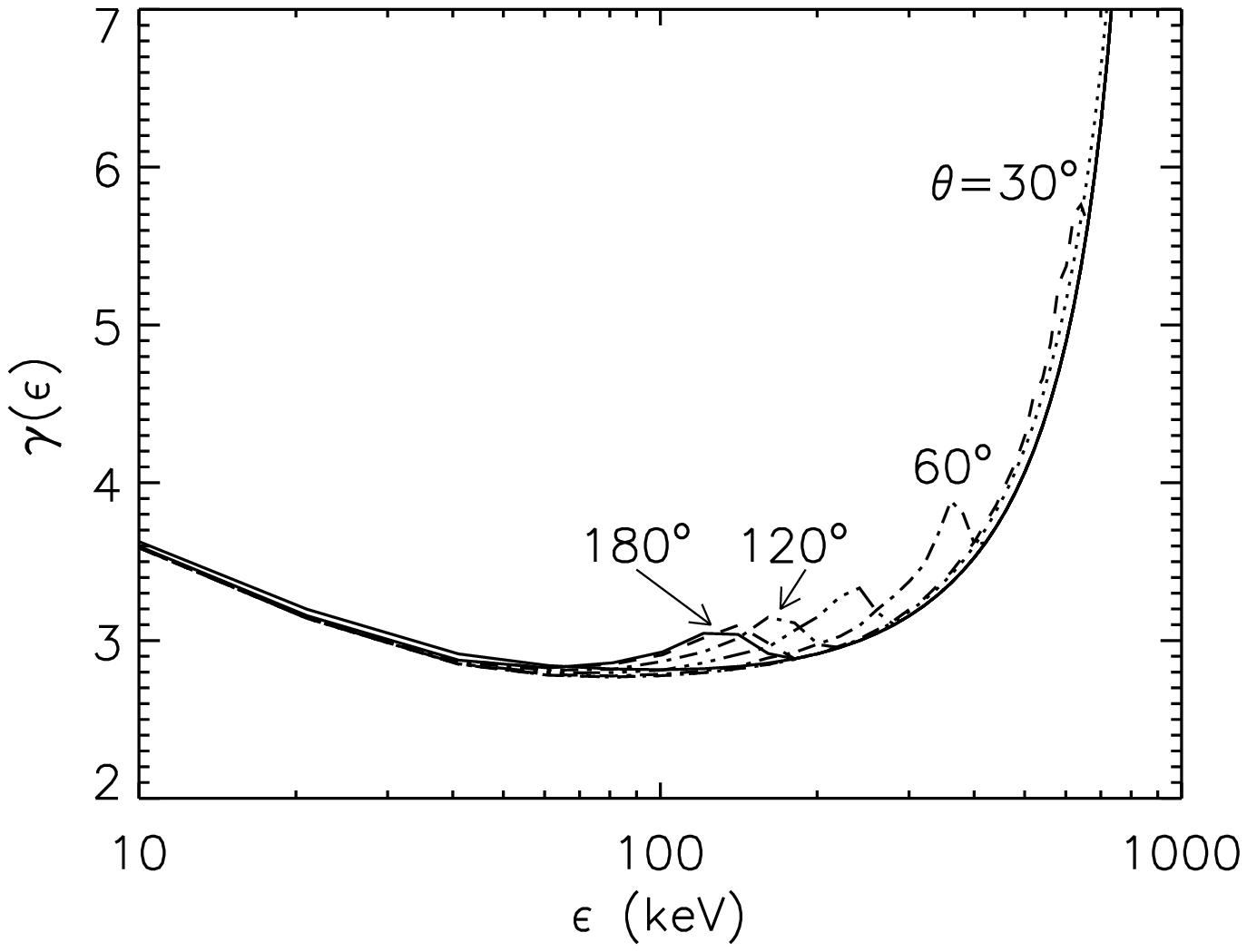}\caption{{\it Left
panel:} Ratio of the total photon spectrum to that produced by
electron-ion bremsstrahlung alone, for four different viewing
angles $\theta$. The mean source electron spectrum is a power law
with spectral index $\delta =2$ and an upper cutoff energy at
1~MeV; {\it Right panel:} Corresponding local spectral indices
$\gamma(\epsilon) = -d \log I(\epsilon)/d \log \epsilon$.  Note
the sharp features in $\gamma$ caused by the absence of
electron-electron emission above a certain photon energy
$\epsilon_{\rm max}$ (see Figure~\ref{epsmax}).  For clarity, only
select curves have been labelled in both panels.} \label{theta}
\end{figure}

Note also that the maximum photon energy $\epsilon _{\rm max}$
depends significantly on the viewing angle $\theta$. Hence, if the
injected electron distribution is highly beamed, the strong
angular dependence of the maximum photon energy produced permits a
determination of the direction of the beam. For example, for
$E_{\rm max}=1$~MeV and $\theta =90^\circ$, there should be
evidence for such a high-energy cutoff in the photon spectrum
around 250~keV (Figure~\ref{epsmax}). Figure~\ref{theta} shows the
effect of such a 1~MeV upper energy cutoff on the total
(electron-ion + electron-electron) photon spectrum $I(\epsilon)$
and on its local spectral index $\gamma = -d \log I(\epsilon)/d
\log \epsilon$. There is an abrupt step in $\gamma$ at $\sim
250$~keV; this step moves towards larger energies as $\theta$ is
reduced, so that the inferred value of $E_{\rm max}$ depends on
the value of $\theta$ appropriate.

\section{Data analysis}

In selecting suitable events for analysis, we searched for a clear
identification of high-energy photons in the flare light curve,
and specifically a count rate high enough to provide good count
statistics in energy channels above 300~keV. Quasi-logarithmic
energy binning was used in order to enhance the signal-to-noise
ratio in each energy channel and the time bins were chosen equal
to RHESSI's rotation period (as given for the time of the flare)
to ensure that there is no differential modulation of the light
curve from varying aspects of the imaging grids.

The data were corrected for the following effects: decimation,
detector energy response, detector livetime, attenuator
transmission, imaging grid transmission, and pulse pile-up.  These
steps were performed using standard software incorporating the
most up to date information on the instrumental calibration
(Schwartz et al. 2002). The background was then subtracted by
using the SPEX package to interpolate between two background time
intervals, one before and one after the flare. Data from detectors
2 or 7 were not used, because their energy resolution is
significantly poorer than for the other detectors (Smith et~al.
2002).

\begin{figure}[pht]
\epsscale{0.99} \plotone{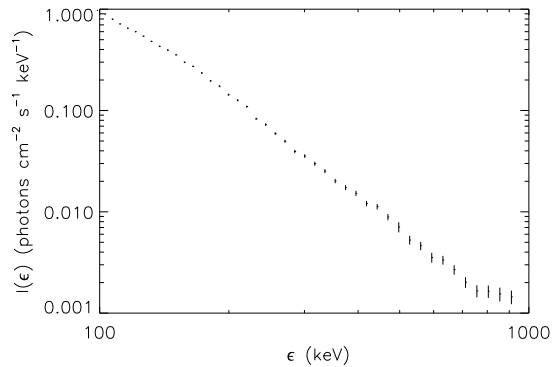} \caption{Photon spectrum for the
time interval 09:43:16 -- 09:44:24 UT in the 2005 January 17 event
with gamma-lines removed.} \label{photon}
\end{figure}

Figure~\ref{photon} shows the photon spectrum for the time
interval 09:43:16 -- 09:44:24 UT (the time of approximate peak
flux) for the 2005 January 17 (GOES class X3.8) event. This event,
which produced several strong gamma-ray lines, was previously
studied by Kontar \& Brown (2006), who concluded that the pitch
angle distribution for electrons up to $\sim 300$~keV is close to
isotropic. We focus attention on the highest energy spectrum
($\epsilon > 200$~keV) in this paper.

This event was located at position $(x=380\arcsec, y=320\arcsec)$
on the solar disk, corresponding to a heliocentric angle $\sim
30^\circ$. Consequently, the assumption of a downward-directed
electron beam leads to angles $\theta$ between the beam direction
and the observer in the second quadrant; this enhances the
possibility of observing the spectral features noted in \S2
associated with the upper limit to electron-electron
bremsstrahlung emission (see Figures~\ref{epsmax} and
\ref{theta}).

\section{Determination of the Mean Source Electron Spectrum}

Before attempting to determine the form of the mean source
electron spectrum responsible for the observed hard
X-ray/gamma-ray continuum, it is first necessary to subtract the
emission from strong gamma-ray spectral lines.  In the energy
range under consideration, the two most significant ranges for
which this subtraction is necessary  are (483-512)~keV and
(829-882)~keV. The corrected spectrum is presented in
Figure~\ref{photon}. The first of these corresponds to the
electron-positron annihilation line at 511~keV and its associated
positronium continuum at lower energies, the second to a variety
of strong emission lines from $^{27}$Al, $^{54}$Cr and $^{56}$Fe
(see Table~1 in Ramaty, Kozlovsky, \& Lingenfelter 1979; Table~1
in Kozlovsky, Murphy, \& Ramaty 2002). These lines were
``removed'' by replacing the data in these ranges with a smooth
interpolation of the continuum spectrum on either side of each
feature.

The residual photon spectra then represent principally
bremsstrahlung continuum, with an emissivity given by
equation~(\ref{def}). These continuum spectra were then used to
determine the mean electron flux spectrum ${\overline F}(E)$ in
the source, using two different, well-established, methodologies
for solution of equation~(\ref{def}).

\subsection{Forward Fit}

Here we follow the procedure of Holman et~al. (2003) and assume
that the mean electron spectrum is the sum of a low-energy
Maxwellian, plus a broken power law of the form

\begin{equation}
{\overline F}(E) = \cases{A \, E^{-\delta_1}; \quad E < E_{\rm
brk} \cr A \, E_{\rm brk}^{\delta_2 - \delta_1} \, E^{-\delta_2};
\quad E \ge E_{\rm brk}.} \label{broken}
\end{equation}
Because the Maxwellian part of ${\overline F}(E)$ (with a
characteristic temperature $T \simeq 3$~keV) is utterly
insignificant at energies $E > 200$~keV, it is not necessary to
consider this component in our analysis.

Calculation, using equation~(\ref{def}), of the photon spectrum
for an ${\overline F}(E)$ of the form~(\ref{broken}), and
comparison with the observed $I(\epsilon)$ above $\epsilon =
200$~keV, permits determination of the best-fit values of the four
parameters $(A, E_{\rm brk}, \delta_1, \delta_2)$.   We performed
such a forward fit for two forms of the bremsstrahlung
cross-section: $Q_{\rm e-i}(\epsilon, E) = Z^2 Q_{e-p} (\epsilon,
E)$ (which includes electron-ion bremsstrahlung only) and $Q_{\rm
tot}(\epsilon, E) = Z^2 Q_{e-p} (\epsilon, E) + Z Q_{e-e}
(\epsilon, E)$ (which includes both electron-ion and
electron-electron bremsstrahlung).  Mean values $<Z> = 1.2$ and
$<Z^2> = 1.44$ (representative of mean solar abundances) were
assumed.

\begin{figure}[pht]
\epsscale{0.99} \plotone{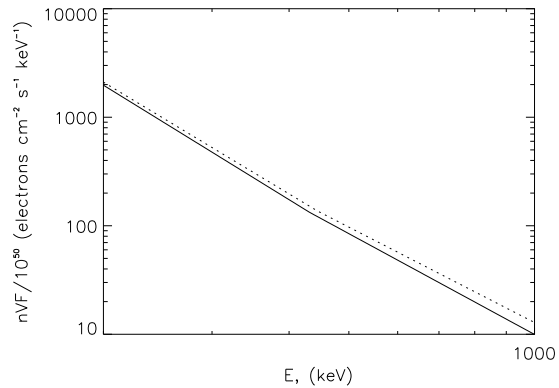} \caption{Forward-fit forms of
$\bar n \, V \, {\overline F}(E)$ for the selected flare. The
dashed curve assumes electron-ion bremsstrahlung only; the solid
curve includes the additional electron-electron term.
\label{ffit}}
\end{figure}

Using the cross-section $Q_{\rm e-i}(\epsilon,E)$, representing
only electron-ion bremsstrahlung, results in best-fit values
$\delta_1 = 3.4,\; \delta_2 = 2.9$, and $E_{\rm brk} = 445 $~keV.
Using the more correct cross-section $Q_{\rm tot}(\epsilon,E)$
(which incorporates both electron-ion and electron-electron terms)
gives $\delta_1 = 3.5, \delta_2 = 3.1 $, and $E_{\rm brk} =
431$~keV. The forms of both of these fits are shown in
Figure~\ref{ffit}. While inclusion of the electron-electron
bremsstrahlung term results in little change to the form of
${\overline F}(E)$ at low energies, its inclusion does lead to the
break energy moving downward from $E \sim 450$~keV to $E \sim
430$~keV, and to the spectral index for the high-energy component
steepening from $\delta \simeq 2.9$ to $\delta \simeq 3.1$
($\Delta \delta \simeq 0.2$). Such a steepening of ${\overline
F}(E)$, and the energy above which it becomes significant, are in
accordance with the expectations expressed in \S1 and with earlier
quantitative estimates based on hardening of hard X-ray spectra
(e.g., Vestrand 1988).

\subsection{Regularized Inversion}

Piana et al. (2003) have demonstrated the construction of smooth, {\it
regularized}, forms for the mean electron flux spectrum
${\overline F}(E)$ from high-resolution {\em RHESSI} photon spectra
$I(\epsilon)$. The advantage of this method is that it is not
necessary to assume an empirical form for the spectrum.
Additionally, as shown by Brown et~al. (2006), this method is
capable of revealing accurately the overall ``shape'' of the electron
spectrum and indicating the presence and approximate form of
small-scale features of sufficient amplitude, if present.

\begin{figure}[pht]
\epsscale{0.99} \plotone{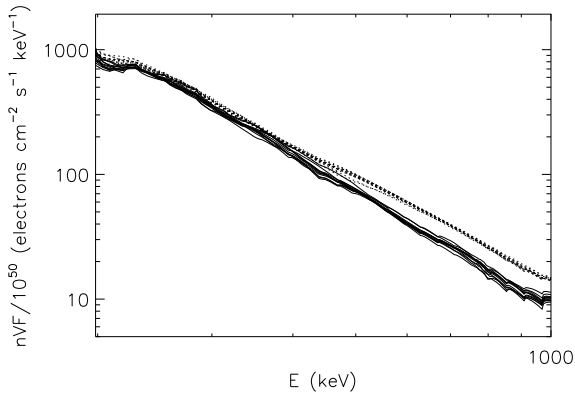} \caption{Recovered forms of the
quantity $\bar n \, V \, {\overline F}(E)$ (in units of
$10^{50}$~electrons~cm$^{-2}$~s$^{-1}$~keV$^{-1}$; see
equation~[\ref{def}]) using a zero-order regularization technique
and presented as a ``confidence strip,'' i.e., a series of
solutions, each based on a realization of the data consistent with
the size of the uncertainties. The dashed lines assume
electron-ion emission only; the solid lines include the additional
electron-electron emission term.} \label{reg_sol}
\end{figure}

Figure \ref{reg_sol} shows the recovered ${\overline F}(E)$
solution for the same photon spectrum used in the forward fit
procedure of Figure~\ref{ffit}.  The results are presented in the
form of a {\it confidence strip}, a set of different realizations
of ${\overline F}(E)$, each curve corresponding to a different
realization of the noisy data set $I(\epsilon)$.

It is clear that the ${\overline F}(E)$ recovered using the full
cross-section~(\ref{qtot}), including electron-electron
bremsstrahlung, is, for $E \gapprox 300$~keV, steeper (spectral
index greater by $\sim 0.4$) than the ${\overline F}(E)$ recovered
assuming purely electron-ion emission. This result is consistent
not only with the forward-fit results of the previous subsection
but also with the physical expectations enunciated in \S1.
Moreover, the dashed confidence strip (corresponding to use of the
electron-ion cross-section alone) has an upward break near
$E=400$~keV (as can be verified visually by looking {\it along} --
rather than {\it at} -- the strip).  However, the true form of
${\overline F}(E)$, as exhibited  by the solid confidence strip,
has a rather featureless power-law form over the energy range from
200 - 1000~keV. Consequently, use of the full cross-section,
including the electron-electron term that becomes important at
energies $\gapprox 300$~keV, removes the need to account for the
$\sim 400$~keV energy that characterizes the (unphysical) upward
break in ${\overline F}(E)$ that appears when only the partial
(electron-ion) cross-section is used in the analysis.

\section{Discussion and Conclusions}

As expected, recognition of the growing importance of
electron-electron bremsstrahlung at high energies reduces, for a
given hard X-ray spectrum, the number of high-energy electrons
required to produce it; this leads to a steepening in the inferred
mean source electron spectrum ${\overline F}(E)$ above $\sim
400$~keV. For the January 17, 2005 event studied, use of the
electron-ion cross-section alone leads, whether by forward fitting
or regularized inversion, to the inference of an upward break
($\Delta \gamma \simeq 0.4$) in ${\overline F}(E)$ at $E \simeq
400$~keV (Figure~\ref{reg_sol}). However, when both electron-ion
and electron-electron bremsstrahlung emission are considered, this
break disappears, resulting in an ${\overline F}(E)$ that has a
straightforward power-law form over the energy range from 100 -
1000~keV. Single-power law suggests electron acceleration without
characteristic energy and corresponding characteristic scale.
Careful interpretation is therefore necessary when faced with
apparent hard X-ray spectral changes in this energy range.

One process that can, for a sufficiently strong magnetic field,
operate strongly in the few 100 keV range and so affect this
argument is gyrosynchrotron emission. However, the presence of
this additional emission mechanism would cause the ${\overline
F}(E)$ to bend {\it downward} at higher energies. The fact that,
after inclusion of the electron-electron contribution, ${\overline
F}(E)$ has no such bend puts an upper limit on the importance of
gyrosynchrotron emission and so an upper limit ($\sim 10$~kG) on
the strength of the ambient magnetic field.

Trottet et~al. (1998) report very significant upward breaks
($\Delta \gamma \simeq 1.2 - 2; E_{\rm brk} \simeq 400$~keV) in
the hard X-ray spectrum for a series of intervals during an
electron-dominated gamma-ray event on 1990 June 11. We agree with
these authors that the inclusion of electron-electron
bremsstrahlung cannot account for such breaks.  However, Vestrand
(1988) reports that ``most flares show a break $\simeq 0.5$''
occurring at an energy ``$\simeq 300 - 400$~keV'' and a similar
statement is made by Dennis (1985) (however, he also reports a
much larger spectral break [$\Delta \gamma \simeq 2$] in an event
observed on June 4, 1980). Such modest ($\Delta \gamma \simeq
0.5$) upward breaks at a photon energies $\epsilon \simeq 300 -
400$~keV are naturally accounted for by including the contribution
from electron-electron bremsstrahlung; other considerations, such
as energy-dependent anisotropy (Li 1995) or a separate
emission/acceleration process (e.g., Heristchi 1986) are in
general not required.

Only features common to all (or at least nearly all) realizations
of ${\overline F}(E)$ can be considered real. Using this
criterion, one must concede that the recovered confidence strip
(Figure~\ref{reg_sol}) is sufficiently wide that no firm evidence
for a sudden change in the local spectral index $\gamma$ (cf.
Figure~\ref{theta}) can be claimed.  Hence the data do not provide
compelling evidence for either strong beaming of the accelerated
electrons or an upper-energy cutoff $E_{\rm max}$ in the
accelerated electron energy distribution.  Such an assessment is
bolstered by Kontar \& Brown's (2006) finding, using a comparison
of the brightness of the primary source with that of the
photospherically-backscattered albedo patch (Kontar et~al. 2006),
that the electron distribution at energies $E \lapprox 200$~keV in
the 2005 January 17 event was also consistent with isotropy.

\acknowledgments EPK was supported by a PPARC Advanced Fellowship
and by a grant from the Royal Society; AGE was supported by NASA
Grant NNG04G063G and by subcontract SA4878-26308 from the
University of California, Berkeley.  The overall effort has
greatly benefited from support by a grant from the International
Space Science Institute (ISSI) in Bern, Switzerland.

%\clearpage %---------------------------------------------------------

\end{document}